\documentclass[a4paper]{article}

\usepackage{INTERSPEECH2022}
\usepackage{comment}
\usepackage{url}
\usepackage{hyperref}
\usepackage{amsmath,amssymb}
\usepackage{subfigure}
\hypersetup{
    colorlinks = false,
}
\usepackage{cite}
\title{DRSpeech: Degradation-Robust Text-to-Speech Synthesis with Frame-Level and Utterance-Level Acoustic Representation Learning}
\name{Takaaki Saeki$^{1,2}$, Kentaro Tachibana$^1$, and Ryuichi Yamamoto$^1$}
\address{$^1$LINE Corporation, Japan. $^2$The University of Tokyo, Japan.}
\email{takaaki\_saeki@ipc.i.u-tokyo.ac.jp, kentaro.tachibana@linecorp.com}

\begin{document}

\maketitle
\begin{abstract}\vspace{-1mm}
Most text-to-speech (TTS) methods use high-quality speech corpora recorded in a well-designed environment, incurring a high cost for data collection.
To solve this problem, existing \textit{noise-robust TTS} methods are intended to use noisy speech corpora as training data. However, they only address either time-invariant or time-variant noises.
We propose a \textit{degradation-robust TTS} method, which can be trained on speech corpora that contain both additive noises and environmental distortions.
It jointly represents the time-variant additive noises with a frame-level encoder and the time-invariant environmental distortions with an utterance-level encoder.
We also propose a regularization method to attain clean environmental embedding that is disentangled from the utterance-dependent information such as linguistic contents and speaker characteristics.
Evaluation results show that our method achieved significantly higher-quality synthetic speech than previous methods in the condition including both additive noise and reverberation.
\end{abstract}

\vspace{-2mm}
\section{Introduction}
\vspace{-1mm}

Text-to-speech (TTS) synthesis is intended to artificially synthesize human speech utterances.
Because TTS models are generally trained on high-quality speech corpora recorded in a well-developed environment, the data collection is time-consuming and financially expensive.
The cost of data collection can be greatly reduced by using data recorded with smartphones~\cite{Raza2018RapidCO} or data downloaded from the Internet~\cite{Takamichi2021JTubeSpeechCO}.
However, those speech data contains distortions such as noise and reverberation; it is challenging to train high-quality TTS models using such data. 
Some studies have been conducted on \textit{noise-robust TTS}: training high-quality TTS models to be robust to time-variant additive noises~\cite{Zhang2021DenoispeechDT} or time-invariant background noises~\cite{hsu2018hierarchical}.
However, real speech data includes both additive noises, which exist independently of the speech signal, and environmental distortions, which is multiplied on the speech signal (e.g., reverberation).
Therefore, there is a need for a TTS framework that jointly represents these types of distortion.

We propose a \textit{degradation-robust TTS} method, which can be trained on degraded speech that contains both time-variant additive noises and time-invariant environmental distortions.
Our method uses a frame-level noise encoder to represent time-variant additive noises and an utterance-level environment encoder to address time-invariant environmental distortions; the two encoders are jointly trained with the TTS model.
For the modeling of environmental distortions, we also propose a regularization method to attain clean environmental embedding that is disentangled from the utterance-dependent information such as linguistic contents and speaker characteristics.
The main contributions of this work are as follows:
\vspace{-1mm}
\begin{itemize} \leftskip -5.5mm \itemsep -0.5mm
    \item  We propose a degradation-robust TTS method, which jointly addresses additive noises and environmental distortions.
    \item We propose a regularization method to acquire clean environmental embedding that is disentangled from linguistic contents and speaker information.
    \item  Evaluation results show that our method outperformed previous methods with respect to objective and subjective metrics in the condition including both noise and reverberation.
\end{itemize}

\vspace{-2mm}
\section{Related work}\label{sec:related}
\vspace{-1mm}

Noise-robust TTS methods, which learn from noisy speech, have been studied.
Some previous studies on noise-robust TTS have used a speech enhancement model; for example, a TTS model has been trained with enhanced speech or features~\cite{ValentiniBotinhao2016InvestigatingRS, valentinibotinhao16interspeech}, and a model trained with clean speakers has been adapted to noisy new speakers~\cite{Dai2020NoiseRT}.
The existing method that is most similar to ours is one in which a TTS model is trained with frame-level noise representation~\cite{Zhang2021DenoispeechDT}.
The aim of our work is to novelly extend that method to represent both time-variant additive noise and time-invariant distortions of degraded speech.
Some previous studies have modeled background noise or distortion~\cite{hsu2018hierarchical,hsu2019conditional,Tan2021EnvironmentAT} without an external speech enhancement model.
However, they only focus on time-invariant noise, limiting the applicability.
Our method uses a style token layer~\cite{wang2018style} for the utterance-level acoustic encoder. 
The original paper~\cite{wang2018style} also reports experiments conducted in the reverberation condition, with the manually selected token representing clean environmental conditions.
In contrast, our method introduces a regularization method to automatically obtain clean environmental embedding.

Neural sequence-to-sequence TTS methods~\cite{wang17tacotron,shen17tacotron2,Ren2019FastSpeechFR,kim2021conditional} that use clean speech corpora have been proposed, including FastSpeech~2~\cite{Ren2021FastSpeech2F}, which is used in our work.
Our method can also be applied to other sequence-to-sequence TTS models~\cite{shen17tacotron2,Kim2017GenerationOL} by conditioning the frame-level and utterance-level embedding on the intermediate features after the duration prediction.

\vspace{-2mm}
\section{Method}\label{sec:method}
\vspace{-1mm}

\begin{figure*}[t]
  \centering
  \includegraphics[width=0.83\linewidth, clip]{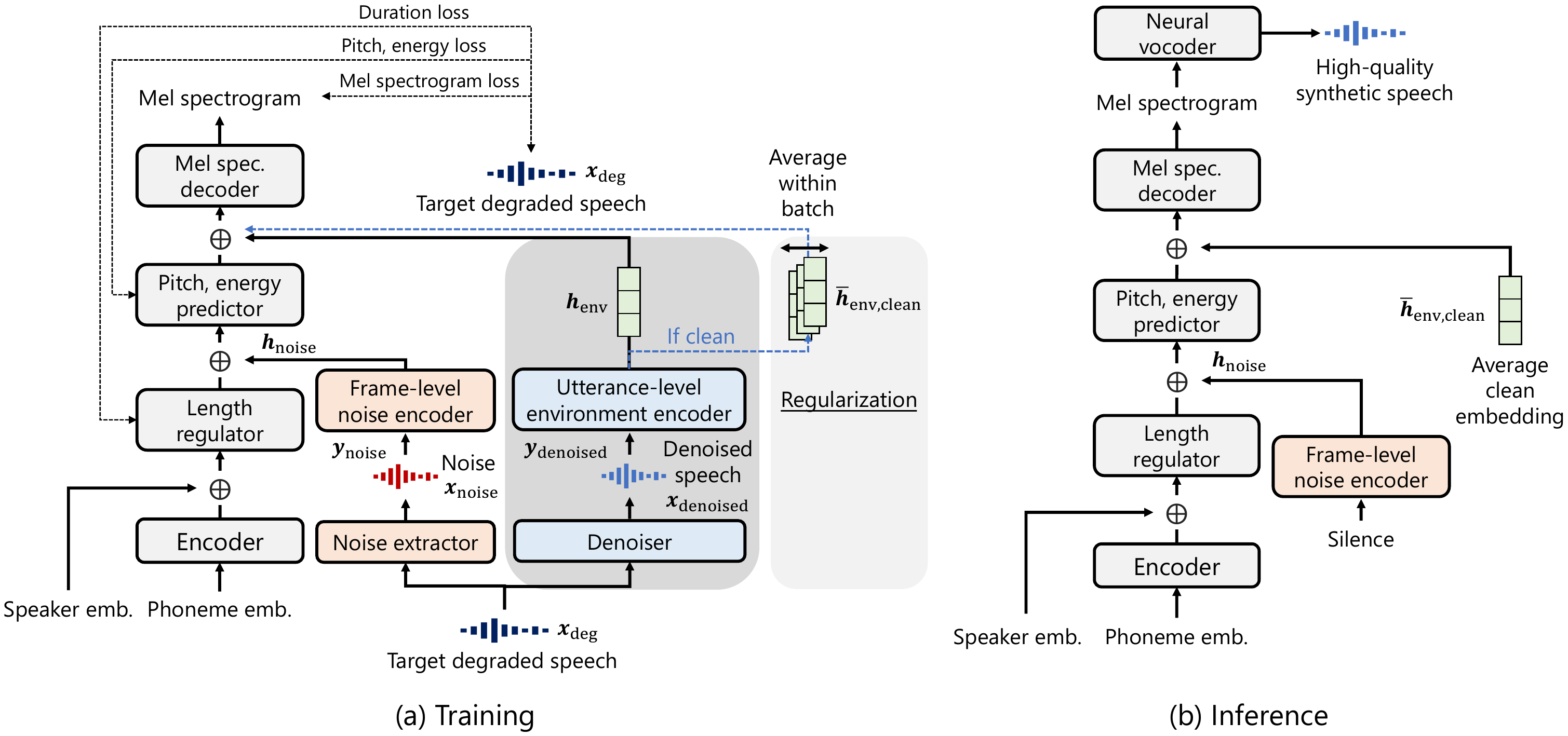}
  \vspace{-2mm}
  \caption{(a) Proposed model is trained on speech corpora that include additive noise and environmental distortions by jointly using frame-level noise representation and utterance-level environmental representation. (b) During inference, model generates high-quality speech by using silence as the input of the noise encoder and average clean environmental embedding as utterance-level representation.}
  \label{fig:method}
  \vspace{-5mm}
\end{figure*}

Fig.~\ref{fig:method} shows an overview of our method.
It jointly uses a frame-level noise representation and an utterance-level environmental representation.
Our model is based on FastSpeech~2~\cite{Ren2021FastSpeech2F}, which consists of a phoneme encoder that encodes the input phoneme embedding into a hidden sequence, a length regulator that predicts and extends the length of the encoded representation, a pitch and energy predictor, and a decoder that outputs acoustic features from an intermediate representation.
The model is a multispeaker model, in which the embedded speaker ID is added to the output of the phoneme encoder.

\vspace{-1mm}
\subsection{Frame-level noise representation learning}\label{sec:method-local}
\vspace{-1mm}
To train the TTS model in a manner that takes time-variant noise into account, we introduce a frame-level noise representation with a noise extractor, as in a previous study~\cite{Zhang2021DenoispeechDT}.
The previous study showed that the TTS model using an intermediate frame-level noise representation achieves higher synthetic speech quality than simply using denoised speech feature as the training target.
Whereas they use a simple U-Net~\cite{Ronneberger2015UNetCN} model as a noise extractor, we use an improved noise extractor in the waveform domain, Conv-TasNet~\cite{Luo2019ConvTasNetSI}, which generalizes to a wider variety of types of additive noise.
The noise extractor extracts only additive noise from the target degraded speech, and a noise encoder outputs a hidden noise representation.

As shown in Fig.~\ref{fig:method}(a), the target degraded waveform $\Vec{x}_{\mathrm{deg}}$ is input to the noise extractor, which is pre-trained to output a noise waveform from a noisy speech waveform with a large dataset.
The output noise waveform $\Vec{x}_{\mathrm{noise}}$ is then converted into a mel spectrogram $\Vec{y}_{\mathrm{noise}}$ and input to the frame-level noise encoder.
The noise encoder outputs a noise representation $\Vec{h}_{\mathrm{noise}}$, which has the same number of frames as the target mel spectrogram and is added to the output of the length regulator.
In inference, we use the silence defined as $\Vec{x}_{\mathrm{noise}}(n) = 0 \, (\text{for all} \, n)$ to generate output speech without frame-level additive noise, as shown in Fig.~\ref{fig:method}(b).

\vspace{-1mm}
\subsection{Utterance-level environmental representation learning}\label{sec:method-global}
\vspace{-1mm}

In addition to a frame-level noise representation, the proposed method also learns an utterance-level environmental representation.
To extract only the environmental conditions from speech---including additive noise and environmental distortion---we introduce a denoiser to remove additive noise.
We use a denoiser based on Conv-TasNet, which is pretrained to output a denoised speech waveform from a noisy speech waveform.
The target degraded waveform $\Vec{x}_{\mathrm{deg}}$ is input to the denoiser, which outputs the denoised speech waveform $\Vec{x}_{\mathrm{denoised}}$.
The mel-spectrogram $\Vec{y}_{\mathrm{denoised}}$ is obtained from $\Vec{x}_{\mathrm{denoised}}$ and input to the utterance-level environment encoder to obtain the environmental representation $\Vec{h}_{\mathrm{env}}$.
We use a style token layer~\cite{wang2018style} for the uttrance-level environment encoder.
The environmental embedding $\Vec{h}_{\mathrm{env}}$ is conditioned on the TTS model as shown in Fig.~\ref{fig:method}(a).

In the original paper proposing the style token layer~\cite{wang2018style}, they report evaluations of the TTS model trained in reverberation condition.
During inference, they manually select a style token corresponding to the clean condition, which requires substantial time and effort.
Therefore, during inference of our method, the average clean embedding $\bar{\Vec{h}}_{\mathrm{env, clean}}$ is conditioned on the TTS model, where $\bar{\Vec{h}}_{\mathrm{env, clean}}$ is obtained by averaging $\Vec{h}_{\mathrm{env}}$ for all the training data recorded in clean environmental conditions.

However, simply providing the TTS model with $\Vec{h}_{\mathrm{env}}$ from each utterance during training and the average clean embedding $\bar{\Vec{h}}_{\mathrm{env, clean}}$ during inference does not guarantee that the desired degradation-robust training can be achieved.
This is because $\Vec{h}_{\mathrm{env}}$ is utterance-dependent and entangled with speaker characteristics.
Therefore, $\bar{\Vec{h}}_{\mathrm{env, clean}}$ does not necessarily represent the clean conditions, and using this embedding can lead to low-quality synthetic speech.
We thus propose the regularization method described in Section~\ref{sec:method-objective} to obtain a disentangled clean environmental embedding $\Vec{h}_{\mathrm{env}}$.

\vspace{-1mm}
\subsection{Training objective with regularization term}\label{sec:method-objective}
\vspace{-1mm}
The training objective of our model consists of mean squared error (MSE) loss for duration, pitch, and energy in addition to L1 loss for mel spectrogram.
The sum of the above loss functions is denoted by $\mathcal{L}_{\mathrm{main}}$, which is the same as the objective used by the original FastSpeech~2~\cite{Ren2021FastSpeech2F}.

Furthermore, we propose a regularization method that uses the average clean environmental embedding during training.
For the regularization, we introduce subtask learning that uses only speech data with clean environmental conditions.
In the subtask learning, the environmental embedding $\Vec{h}_{\mathrm{env, clean}}$ estimated from target speech is averaged within the batch.
The averaged embedding $\bar{\Vec{h}}_{\mathrm{env, clean}}$ is conditioned on the TTS model as described in Section~\ref{sec:method-global} and the loss function is calculated in the same manner as in the main task.
Let $\mathcal{L}_{\mathrm{average}}$ be the regularization term defined by the subtask learning.
The overall training objective is then defined as:
\begin{equation}
    \mathcal{L} = \mathcal{L}_{\mathrm{main}} + \alpha \mathcal{L}_{\mathrm{average}},
\end{equation}
where $\alpha$ is a weighting term and set to $1.0$ in our evaluations.
With this regularization, the utterance-level encoder can extract only the acoustic aspects shared across multiple utterances (i.e., time-invariant environmental conditions), which are disentangled from the utterance-dependent information such as linguistic contents and speaker characteristics.
The dark gray area in Fig.~\ref{fig:method}(a) depicts the main-task learning to define $\mathcal{L}_{\mathrm{main}}$, and the light gray area depicts the subtask learning to define $\mathcal{L}_{\mathrm{average}}$.

\vspace{-2mm}
\section{Experimental evaluation}\label{sec:evaluation}

\begin{table*}[tb]
\centering
\caption{Results of objective evaluation for all methods}
\vspace{-3mm}
\label{tab:objective}
\scalebox{0.96}{
\footnotesize
\begin{tabular}{l|cccccccc}
\toprule
                      & \multicolumn{2}{c}{Clean} & \multicolumn{2}{c}{Noise} & \multicolumn{2}{c}{Reverb} & \multicolumn{2}{c}{Noise+Reverb} \\ \cmidrule(lr){2-3} \cmidrule(lr){4-5} \cmidrule(lr){6-7} \cmidrule(lr){8-9}
                      & MCD     & Log F0 RMSE     & MCD     & Log F0 RMSE     & MCD      & Log F0 RMSE     & MCD          & Log F0 RMSE         \\ \midrule
Enhancement TTS~\cite{valentinibotinhao16interspeech} & $9.04$  &  $10.12$ &  $9.16$  & $9.86$  &  $9.41$  & $10.34$ &  $10.39$    & $10.33$ \\
Noise-robust TTS~\cite{Zhang2021DenoispeechDT} & $\textbf{7.74}$ & $\textbf{8.00}$ &  $\textbf{7.99}$   &  $\textbf{8.56}$  & $8.64$ &  $8.51$  & $9.65$ &  $9.58$  \\
DRSpeech              & $7.94$ & $8.24$ &  $8.19$  & $8.79$ & $\textbf{8.25}$ & $\textbf{8.09}$   & $\textbf{9.18}$ & $\textbf{9.11}$     \\ \bottomrule
\end{tabular}
}
\vspace{-3mm}
\end{table*}

\vspace{-1mm}
\subsection{Experimental condition}
\vspace{-1mm}

\subsubsection{Database and preprocessing}
\vspace{-1mm}
We used VCTK~\cite{Yamagishi2019CSTRVC}, a multi-speaker English speech corpus, for the degraded speech dataset.
The sampling frequency was set to 22.05~kHz.
As in the previous study~\cite{Zhang2021DenoispeechDT}, we used PNL 100 Nonspeech Sounds~\cite{hu2010tandem} for the noise dataset to generate degraded speech data.
We chose reverberation in this study as the type of environmental distortion for degraded speech.
In previous studies on noise-robust TTS~\cite{hsu2019conditional,Zhang2021DenoispeechDT}, VCTK speakers were divided into two subsets: one for the clean condition and the other for the noise condition.
In this study, in order to consider both noise and environmental conditions, the speakers were randomly divided into four sets: 1) Clean, 2) Noise, 3) Reverb, and 4) Noise+Reverb conditions.
For the Clean set, we used the original VCTK speech data.
In the Noise set, a noise clip was randomly selected from the noise dataset and added to each speech utterance, where the lengths of the noise and speech waveform were matched by repeating or truncating the noise clip.
We set the power of the noise signal according to the loudness units relative to full scale (LUFS)~\cite{recommendation2015lufs}, with the LUFS sampled randomly according to a uniform distribution between -40 and -32.
In the Reverb set, we simulated reverberation using an open-source simulator~\cite{Scheibler2018PyroomacousticsAP}.
We created a room with size $(l, w, h) = (10~$m$, 7.5~$m$, 3.5~$m$)$ and placed a speech source at $(5~$m$, 3~$m$, 1.6~$m$)$ and a monaural microphone at $(0.5~$m$, 4.0~$m$, 0.5~$m$)$, assuming that the speaker was standing in the room.
The $T_{60}$ was set to 0.2~s, which was between the values corresponding to the weak and strong reverberation conditions in the previous study~\cite{Kim2017GenerationOL}.
In the Noise+Reverb set, we placed a noise source at $(3~$m$, 7~$m$, 0.2~$m$)$ and set the other room configuration parameters as for the Reverb set.
We refer to the above dataset as VCTK-degraded.

We randomly selected 512 utterances from VCTK-degraded as the validation set, 512 utterances as the test set, and used the remainder as the training set.
When analyzing speech features, the 80-dimensional mel spectrogram was extracted with a frame size of 1024 and a frame shift of 256. 
The pitch was calculated by the Harvest~\cite{Morise2017HarvestAH} algorithm, and the duration of the phoneme sequence was extracted using Montreal Forced Aligner~\cite{McAuliffe2017MontrealFA}.

We also created a large-scale noisy speech dataset to train the denoiser and noise extractor.
Our dataset was inspired by LibriMix~\cite{Cosentino2020LibriMixAO}, a large-scale dataset for  speech separation based on LibriSpeech~\cite{librispeech}.
LibriSpeech itself is not suitable for TTS because of its low sampling frequency and relatively low speech quality.
Therefore, we created a new corpus named LibriTTSMix, which is based on LibriTTS~\cite{zen19_interspeech} and the wham! noise dataset~\cite{Wichern2019WHAM} used in LibriMix.
We randomly selected a noise clip from the noise dataset and added it to each speech utterance of train-clean-360 and train-clean-100, setting the noise power was set according to a uniform distribution between -38 and -30 LUFS.
The sampling frequency of the datasets was set to 22.05~kHz.
We trained the denoiser to estimate the clean speech from the noisy speech on the dataset, and we trained the noise extractor to estimate the noise signal.

\vspace{-1mm}
\subsubsection{Model details}
\vspace{-1mm}
We used TTS models based on FastSpeech~2~\cite{Ren2021FastSpeech2F}, which has four and six Transformer blocks in the encoder and decoder, respectively.
In each Transformer block, the hidden size of the self-attention and feedforward layers was set to 256, with 256-dimension phoneme embedding.
The dimension of speaker embedding was set to 256, and the number of speakers was 109 when training the model with VCTK-degraded.
For the neural vocoder to generate synthetic speech waveforms from mel spectrograms, we used a pretrained multi-speaker HiFi-GAN\footnote{\scriptsize{\url{https://github.com/jik876/hifi-gan}}}.

We used Conv-TasNet for the noise extractor and denoiser.
We set the number of repeats to two and we followed the published recipe to set the remaining parameters\footnote{\scriptsize{\url{https://github.com/asteroid-team/asteroid/blob/v0.5.3/egs/librimix/ConvTasNet/local/conf.yml}}}.
Because the model architecture of the frame-level noise encoder was not described in the previous paper~\cite{Zhang2021DenoispeechDT}, we set up a new model based on 1D convolution layers.
The model consists of four residual convolution blocks, each of which consists of two 1D-convolution layers with a kernel size of 3, two batch normalization~\cite{iofee15batchnorm} layers,  and a skip connection~\cite{ResNet}.
We used a style token layer~\cite{wang2018style} for the model architecture of the utterance-level environment encoder.
We set the number of heads, the number of tokens, and the dimension of tokens to 8, 10, and 256, respectively.

\vspace{-1mm}
\subsubsection{Training settings}
\vspace{-1mm}
We firstly trained the noise extractor and noise encoder for 300k steps with a batch size of 8.
We used the Adam optimizer~\cite{kingma2014adam} with a learning rate of 0.001, and used the loss function described in the original paper~\cite{Luo2019ConvTasNetSI}.
After pretraining FastSpeech~2 on LibriTTS, we fixed the weights of only the phoneme encoder and used the weights for the initialization of the proposed model and the other baseline models described in Section~\ref{sec:evaluation-results-mos}.
We trained DRSpeech and the other baseline models on VCTK-degraded with a batchsize of 16, which was the maximum batchsize available with a NVIDIA V100 GPU (16~GiB).
During training, the silence was provided as the input of the noise encoder in the Clean and Reverb conditions.
We used the model with the lowest validation mel spectrogram loss for evaluations.
In contrast to the previous study~\cite{Zhang2021DenoispeechDT}, we did not use the ground-truth noise signal, to assume a more realistic situation.

\vspace{-1mm}
\subsection{Results}

\begin{table}[tb]
    \centering
    \setlength{\tabcolsep}{1mm} 
    \caption{Results of MOS evaluation test for all methods}
    \vspace{-3mm}
    \label{tab:mos}
    \subtable[\textbf{Without reverberation}]{
        \footnotesize
        \begin{tabular}{l|cc}
        \toprule
        & \multicolumn{2}{c}{MOS}  \\ \cmidrule(lr){2-2} \cmidrule(lr){3-3}
        & Clean & Noise \\ \midrule
        Clean GT & $4.14 \pm 0.14$ & $4.77 \pm 0.08$   \\
        Degraded GT & - & $2.78 \pm 0.18$  \\ \midrule
        Enhancement TTS~\cite{valentinibotinhao16interspeech} & $3.62 \pm 0.16$ & $2.85 \pm 0.16$     \\
        Noise-robust TTS~\cite{Zhang2021DenoispeechDT} & $3.53 \pm 0.16$  & $2.88 \pm 0.15$ \\ 
        DRSpeech & $\textbf{3.77} \pm 0.15$  & $\textbf{3.30} \pm 0.15$ \\ 
        \bottomrule
        \end{tabular}
    }
    \subtable[\textbf{With reverberation}]{
        \vspace{-4mm}
        \footnotesize
        \begin{tabular}{l|cc}
        \toprule
        & \multicolumn{2}{c}{MOS}  \\ \cmidrule(lr){2-2} \cmidrule(lr){3-3}
        & Reverb & Noise+Reverb \\ \midrule
        Clean GT & $4.62 \pm 0.11$ & $4.71 \pm 0.09$ \\
        Degraded GT & $3.24 \pm 0.17$ & $2.03 \pm 0.15$ \\ \midrule
        Enhancement TTS~\cite{valentinibotinhao16interspeech} & $2.39 \pm 0.15$ & $2.05 \pm 0.15$ \\
        Noise-robust TTS~\cite{Zhang2021DenoispeechDT} & $2.47 \pm 0.16$ & $2.23 \pm 0.16$ \\ 
        DRSpeech & $\textbf{2.80} \pm 0.16$ & $\textbf{2.56} \pm 0.16$ \\ 
        \bottomrule
        \end{tabular}
    }
    \vspace{-5mm}
\end{table}

\vspace{-1mm}
\subsubsection{Evaluation of DRSpeech and baseline methods}\label{sec:evaluation-results-mos}
\vspace{-1mm}
We conducted objective and subjective evaluations to compare DRSpeech with several baseline methods.
We prepared a baseline method named Enhancement TTS, in which VCTK-degraded was processed with the speech enhancement model based on Conv-TasNet and then the FastSpeech~2 was trained on the enhanced data.
This method is a straightforward approach that corresponds to the previous work~\cite{valentinibotinhao16interspeech}.
We also prepared a model that considers only the frame-level noise representation described in Section~\ref{sec:method-local}. This method corresponds to the previous study~\cite{Zhang2021DenoispeechDT} and is named Noise-robust TTS.
It should be noted that this method is not identical to the previous method~\cite{Zhang2021DenoispeechDT}: Noise-robust TTS uses a more powerful noise extractor, does not use an adversarial connectionist temporal classification loss, and fixes the model parameter of the noise extractor.
For a fair evaluation of the proposed method, we kept all conditions the same as those of the proposed method, with the exception of the utterance-level environment encoder.
We refer to the proposed method---using all the methods described in Section~\ref{sec:method}---as DRSpeech.
We also compared the natural speech of VCTK (Clean GT) and the target degraded speech of VCTK-degraded (Degraded GT), used as training data for DRSpeech.
All the TTS methods were trained on the entire training set of VCTK-degraded and evaluated separately for the four conditions (Clean, Noise, Reverb, and Noise+Reverb) in the test set.
Audio samples are publicly available\footnote{\scriptsize{\url{https://takaaki-saeki.github.io/drspeech_demo}}}.

We first conducted objective evaluations.
Two metrics---mel cepstral distortion (MCD)~\cite{fukada92melcep} and log F0 root mean squared error (RMSE)---were calculated for all the TTS methods.
Table~\ref{tab:objective} lists the results.
We observe that Noise-robust TTS and DRSpeech, which learn to take acoustic representation into account, performed better than Enhancement TTS in all conditions.
In addition, the MCD and log F0 RMSE of Noise-robust TTS are both about $0.2$ less than those of DRSpeech in the Clean and Noise conditions.
In the Reverb and Noise+Reverb conditions, DRSpeech achieved scores of $0.4--0.5$ better than those of Noise-robust TTS.

Second, we conducted a mean opinion score (MOS) test to evaluate the naturalness of synthetic speech.
Nine utterances, all by different speakers, were sampled from each of the four degradation conditions in the test set.
Twenty evaluators rated the naturalness on a scale of 1 to 5.
Table~\ref{tab:mos} lists the results.
DRSpeech achieved the highest MOS in all the conditions (Clean, Noise, Reverb, and Noise+Reverb).
In particular, in the Noise+Reverb condition, DRSpeech achieved a MOS of 0.5 greater than that of Enhancement TTS.
In the Noise condition, the average MOS of Noise-robust TTS was greater than that of Degraded GT and Enhancement TTS; this is consistent with the result of the previous study~\cite{Zhang2021DenoispeechDT}.
Moreover, DRSpeech outperformed Noise-robust TTS in the Noise condition.
Because a noise is added to the entire speech utterance, DRSpeech, which considers global conditions, may be expected to achieve better results than Noise-robust TTS.

\begin{table}[tb]
    \centering
    \caption{CMOS of DRSpeech with and without regularization}
    \vspace{-3mm}
    \label{tab:cmos}
    \scalebox{0.9}{
    \footnotesize
    \begin{tabular}{l|cccc}
    \toprule
    & \multicolumn{4}{c}{CMOS}  \\ \cmidrule(lr){2-2} \cmidrule(lr){3-3} \cmidrule(lr){4-4} \cmidrule(lr){5-5}
    & Clean & Noise & Reverb & Noise+Reverb \\ \midrule
    DRSpeech &  0 & 0   & 0  & 0     \\
    w/o Regularization &  $-0.207$ & $-0.177$ & $-0.205$ & $0.005$ \\ \bottomrule
    \end{tabular}
    }
    \vspace{-4mm}
\end{table}

\vspace{-1mm}
\subsubsection{Evaluation of regularization}\label{sec:evaluation-results-cmos}
\vspace{-1mm}
We conducted a CMOS (Comparison MOS) test to evaluate the effectiveness of the regularization method separately from the utterance-level environment encoder.
In this evaluation, two synthetic speech samples from DRSpeech---with and without the regularization method---were played in the random order.
Twenty evaluators rated the naturalness of the second speech utterance relative to the first one on a 7-point scale from -3 to +3.
Ten pairs with the same utterance text were sampled from the test sets, all with different speakers.
Table~\ref{tab:cmos} lists the results.

DRSpeech with the regularization outperformed DRSpeech without the regularization by about 0.2 in the Clean, Noise, and Reverb conditions.
However, there was only 0.005 difference between the two methods in the Noise+Reverb condition.

\vspace{-1mm}
\subsection{Analysis}
\vspace{-1mm}
In the objective evaluation, Noise-robust TTS outperformed DRSpeech under the Clean and Noise conditions.
In the subjective evaluation of Section~\ref{sec:evaluation-results-mos}, however, DRSpeech outperformed Noise-robust TTS in both conditions.
A discrepancy between objective and subjective evaluation results has often been observed in other studies~\cite{Weiss2021WaveTacotronSE,Hayashi2021ESPnet2TTSET}.
From the standpoint that subjective metrics are more reliable than objective metrics, we can conclude that DRSpeech achieves higher-quality synthetic speech than the Noise-robust TTS.
Furthermore, DRSpeech outperformed the baseline methods in objective and subjective evaluations in the Reverb and Noise+Reverb conditions, indicating the effectiveness of DRSpeech in the reverberant conditions.

The evaluation of Section~\ref{sec:evaluation-results-cmos} confirms the effectiveness of the regularization method in the Clean, Noise, and Reverb conditions, although there was no significant difference in the Noise+Reverb condition.
This means that the improvement achieved by the utterance-level encoder was more dominant than that achieved by the regularization method in the presence of both noise and reverberation.
Further improvement of the regularization method is required in future research.

\vspace{-2mm}
\section{Conclusions}
\vspace{-1mm}
We proposed a degradation-robust TTS method with a frame-level noise representation and an utterance-level environmental representation.
Our method also uses a regularization method to obtain utterance-independent clean environmental embedding during training.
Experimental results showed that our method outperformed existing noise-robust TTS methods with respect to both objective and subjective measures in conditions including reverberation. 
Our future work includes conducting experiments with a wider variety of types of distortion. 

%\section{Acknowledgements}

\bibliographystyle{IEEEtran}
\bibliography{tts}

\end{document}